\documentstyle[preprint,aps,epsfig,amstex]{revtex}
\begin{document}
\draft
\title{The Hartree approximation in dynamics of polymeric manifolds in the melt}
\author{V.G.Rostiashvili$^{(a,b)}$, M. Rehkopf$^{(a)}$ and
        T.A. Vilgis$^{(a)}$}
\address{$^{(a)}$Max-Planck-Institut f\"ur Polymerforschung, Postfach 3148,
D-55021
  Mainz, Germany}
\address{$^{(b)}$ Institute of Chemical Physics, Russian Academy of Science,
142432, Chernogolovka, Moscow region, Russia}
\date{\today}
\maketitle
\begin{abstract}
The Martin-Siggia-Rose (MSR) functional integral technique is applied to the
dynamics of a D - dimensional manifold in a melt of similar manifolds. The integration over
the collective variables of the melt can be simply implemented in the
framework of the dynamical random phase approximation (RPA). The resulting
effective action functional of the test manifold is treated by making use of
the selfconsistent Hartree approximation. As
an outcome the generalized Rouse equation (GRE) of the test manifold is
derived and its static and dynamic properties are studied. It was found that
the static upper critical dimension, $d_{\rm uc}=2D/(2-D)$, discriminates between
Gaussian 
(or screened) and non-Gaussian regimes, whereas its dynamical counterpart,
${\tilde d}_{uc}=2d_{\rm uc}$, distinguishes between the simple Rouse and
the renormalized Rouse behavior. We have argued that the Rouse mode correlation
function has a stretched exponential form.
The subdiffusional exponents for this regime are calculated explicitly. The
special case of linear chains, $D=1$, shows good agreement with
Monte-Carlo-simulations (MC).
\end{abstract}
\section{Introduction}
There has been recent  interest in the dynamical behavior of polymers
(or more generally polymeric manifolds)
in a quenched disordered random medium \cite{1,2}
and in a melt \cite{3,4,4',4'',5,6}. The dynamics of flux-lines in a type II
superconductor \cite{7} also belongs to the same class of problems as in
\cite{1,2}. The derivation of the equations of motion for the time correlation
functions has been carried out
by making use either of the projection formalism and
the mode-coupling approximation \cite{3,4,4',4'',5} or the Martin-Siggia-Rose
(MSR) functional technique and the selfconsistent Hartree approximation \cite{1,6,7}.
The latter theoretical  approach (which is equivalent to the Hartree type
approximation) was earlier also successfully applied to the investigation of
static properties of different models with \cite{8,9,10} or without
\cite{11,12,13} replica symmetry breaking.

The basic description for polymer dynamics in general is the so-called
Rouse model \cite{14,deGennes}, where the polymer configuration is expressed in dynamical
modes. The physical background of the Rouse model is very simple: It
corresponds to a non interacting chain stirred by a white noise random force,
in the usual Langevin sense.
It is well known from experiments
\cite{Richter},
but also very surprising
that this  Rouse model provides a good
description for the melt of the relatively short chains $N<N_{e}$. At higher
degrees of polymerization the
dynamics of the long chain can be described by the reptation model
\cite{14,deGennes,15}.
For the case of the short chain melt,
it is  not obvious that the collisions of the surrounding chains
close to a test chain add  up to a white noise force. On the other hand,
the obvious thought to explain the Rouseian behavior in short chain polymer
melts is first the excluded volume screening and second the inactivity of
topological constraints ("entanglements") at these length scales.
The chain in a melt has  Gaussian statistics, i.e., $R \propto \sqrt{N}$ can be
explained by the screening of the excluded volume interactions \cite{14,deGennes}. In
this case the chains (or generally speaking polymeric manifolds) strongly
interpenetrate. The screening of exclude volume reduces the upper critical
dimension of the interactions (from four to approximately
two in the case of linear chains), and we can expect that for higher connected
polymers such as $D$ - dimensional manifolds the screening becomes also
dependent upon the embedding space dimension $d$ and the connectivity.

Nevertheless we show below in this paper that
in the case of  dynamics the situation is more
complicated, even for linear chains.
The screening of the excluded volume interactions leads by no
means automatically to the Rouse dynamics even for the short chains.
Indeed we will show that the interactions introduce a new dynamical regime
in $2 \leq d \leq 4$. This new regime is derived on different grounds than those
proposed by Schweizer \cite{4,4'}. Moreover we need to resolve the following questions:
How does
the bare monomeric friction coefficient $\xi_{0}$ renormalize due to the
interactions of the test chain with the surrounding melt? Under which
conditions is renormalization relevant?
A short presentation of the results of this paper
was given in a recent letter \cite{17}.

We use here the MSR-functional technique and the selfconsistent Hartree
approximation for the
investigation of the static and mainly dynamic properties of a polymeric $D$ -
dimensional manifold \cite {18} (or a fractal) in the melt of the same
species. One of the main results of the present study is the derivation of a
generalized Rouse equation (GRE) for such D-dimensional manifold. In this
equation the static and dynamic parts are treated on a equal footing and in
the static limit the screening and saturation of $D$ - dimensional manifolds
are reproduced in a different way than in \cite{19}.

We should stress that the manifolds in our consideration have phantom springs,
i.e. are crossable, so that entanglements cannot occur and the reptational
dynamics is not considered here. The reptational dynamics is driven by
topological constraints and will be considered in a subsequent publication. We
describe below the manifolds only in terms of connectivity and excluded volume
interactions. The connectivity defines the $D$ - dimensional subspace which is
embedded in the Euclidean space of $d$ dimensions.

This model includes the cases between linear polymer chains, which correspond
to $D=1$, and tethered membranes (D=2). By analytic continuation to rational
numbers of the spectral dimension statements on polymeric fractals can be
made. Branched polymers (and percolation clusters) correspond closely to the
spectral dimension $D=4/3$. In a series of papers
\cite{19} one of us has considered the different regimes in static
scaling. Here we will find that these regimes besides the Gaussian one are
unstable. Nevertheless we will show below that a new dynamical regime for the
motion of manifold segments appears. The whole dynamical consideration results
in a sub-diffusive behavior and exponents, which are confirmed for a melt of
linear chains in 3 -  dimensions by Monte-Carlo (MC) numerical investigations
\cite{Paul,Smith,Kopf,Paul1,Paul2} (see also \cite{20,20',21}).

The paper is organized as follows. In Sec.~II, we define the system and derive
the GRE. To do this we integrate over the matrix collective variables. The
resulting action in terms of the test manifold variables will be treated in
the framework of the Hartree approximation. In Sec.~III, on the basis
of the GRE the static and dynamic properties of the test manifold in the melt
are investigated systematically. We discuss the concept of the static and
dynamic upper critical dimensions and calculate explicitly the dynamical
exponents. Discussions and conclusions can be found in Sec.~IV. In the light  
of the GRE some
criticism of the Polymer Mode Coupling Approximation (PMCA) by Schweizer \cite{4,4',4'',5} is
relegated to Appendix A.
\section{Derivation of the GRE}
\subsection{Integration over the collective variables}
We consider a melt of $D$ - dimensional manifolds which is embedded in the $d$
- dimensional space. The test manifold is represented by the $d$ - dimensional
vector ${\bf R}({\vec x},t)$ with the $D$ - dimensional vector ${\vec x}$ of
the internal coordinates which labels the beads. The total number of beads is
given by ${\cal N}=N\times N\times \ldots\times N $ ($D$ times). In the same way the manifolds which belong to the
surrounding matrix are characterized by ${\bf r}^{(p)}({\vec x},t)$
$(p=1,2,\ldots ,M)$. The notations are taken in such a way, that the boldfaced
characters describe the external degrees of freedom in the $d$ - dimensional
Euclidean space, whereas the arrow hatted vectors correspond to the internal
$D$ - dimensional space. The model of the melt of $M$ (monodisperse) tethered
manifolds used in the following is based on the generalized Edwards
Hamiltonian
\begin{eqnarray}
H=\frac{1}{2}\epsilon\sum_{p=1}^{M}\int d^{D}x\left(\nabla_{{\vec x}}{\bf r}^{(p)}({\vec
  x})\right)^{2}+\frac{1}{2}\sum_{p,p'=1}^{M}\int d^{D}x
\int d^{D}x' V\left[{\bf r}^{(p)}({\vec x})-{\bf r}^{(p')}({\vec
    x}')\right]\label{Hamilton}
\end{eqnarray}
where $\epsilon=dT/l^{2}$ is the elastic modulus with the Kuhn segment length
$l$ and we consider units defined such that the Boltzmann constant $k_{B}=1$.

An additional test manifold is immersed in this melt and is described by the
variables ${\bf R}({\vec x},t)$. As a result the Langevin equations in
Cartesian components $j$ for the whole system have the form
\begin{eqnarray}
\xi_{0}\frac{\partial}{\partial
t}R_{j}(\vec{x},t)&-&\varepsilon\Delta_{x}R_{j}(\vec{x},t)\nonumber\\
&+&{\frac{\delta}{\delta R_{j}(\vec{x},t)}\int
d^{D}x' V\left({\bf R}(\vec{x},t)-{\bf
R}(\vec{x}',t)\right)}\nonumber\\&+&{\frac{\delta}{\delta R_{j}(\vec{x},t)}\sum_{p=1}^{M}\int
d^{D}x' V\left({\bf R}(\vec{x},t)-{\bf
r}^{(p)}(\vec{x}',t)\right)}=f_{j}(\vec{x},t)\label{R}
\end{eqnarray}
and
\begin{eqnarray}
\xi_{0}\frac{\partial}{\partial
t}r_{j}^{(p)}(\vec{x},t)&-&\varepsilon\Delta_{x}r_{j}^{(p)}(\vec{x},t)\nonumber\\&+&\frac{\delta}{\delta
r_j^{(p)}(\vec{x},t)}\sum_{m=1}^M\int d^{D}x'V\left({\bf r}^{(p)}(\vec{x},t)-{\bf
r}^{(m)}(\vec{x}',t)\right)\nonumber\\&+&\frac{\delta}{\delta
r_j^{(p)}(\vec{x},t)}\int d^{D}x' V\left({\bf r}^{(p)}(\vec{x},t)-{\bf
R}(\vec{x}',t)\right)={\tilde f}_j(s,t)\label{r}
\end{eqnarray}
where $\xi_{0}$ is the bare friction coefficient, $V(\cdots)$ is the excluded
volume interaction function, $\Delta_{{\vec x}}$ denotes a $D$ - dimensional
Laplacian in internal space and the random forces have the standard Gaussian distribution.

We find it more convenient to reformulate the Langevin problem
(\ref{R})-(\ref{r}) in the MSR-functional integral representation \cite{6}. The
generating functional (GF) of this problem can be written as 
\begin{eqnarray}
Z\left\{\cdots\right\}&=&\int DR_j(\vec{x},t)D{\hat
  R}_j(\vec{x},t)\nonumber\\
&\times&\exp\left\{-\Xi\left[R_j(\vec{x},t),{\hat
  R}_j(\vec{x},t)\right]-A_{1}\left[{\bf R}(\vec{x},t),\hat{\bf
  R}(\vec{x},t)\right]\right\}\label{GF1}
\end{eqnarray} 
where dots imply some source fields and the action functional of the test
manifolds is given by
\begin{eqnarray}
A_{1}\left[{\bf R}({\vec x},t),{\hat {\bf R}}({\vec x},t)\right]=&-&\int
d^{D}x\int dt\Bigg\{i{\hat R}_{j}({\vec
  x},t)\left[\xi_{0}\frac{\partial}{\partial t}R_{j}({\vec
  x},t)-\epsilon\Delta_{x}R_{j}({\vec
  x},t)\right]\nonumber\\
&-&\int d^{D}x' i{\hat R}_{j}({\vec
  x},t)\frac{\delta}{\delta R_{j}({\vec
  x},t)}V\left[{\bf R}(\vec{x},t)-{\bf R}(\vec{x}',t)\right]\nonumber\\&+&T\xi_{0}\left[i{\hat R}_{j}({\vec
  x},t)\right]^{2}\Bigg\}\label{testaction}.
\end{eqnarray}  
Here the hatted vectors ${\hat {\bf R}}$ describe the auxilliary (response)
fields in the standard MSR technique. Their correlation functions are given by
$<{\hat R}_{j}{\hat R}_{l}>=0$ (Causality) and $i<{\hat R}_{j}R_{l}>$ is a
response function. 

The influence functional in eq.~(\ref{GF1}) has the form
\begin{eqnarray}
\Xi\left[{\bf R},{\bf \hat
 { R}}\right]=&-&\ln\int\prod_{p=1}^{M}D{\bf r}^{(p)}(\vec{x},t)D\hat{\bf r}^{(p)}(\vec{x},t)\nonumber\\
&\times&\exp\Bigg\{-A_{2}\left[{\bf r}^{(p)},\hat{\bf r}^{(p)}\right]\nonumber\\
&+&\sum_{p=1}^M\int d^{D}xd^{D}x'\int dt
i\hat{R}_{j}(\vec{x},t)\frac{\delta}{\delta R_{j}(\vec{x},t)}V\left({\bf R}(\vec{x},t)-{\bf
r}^{(p)}(\vec{x}',t)\right)\nonumber\\
&+&\sum_{p=1}^M\int d^{D}xd^{D}x'\int dt
i\hat{r}_{j}^{(p)}(\vec{x},t)\frac{\delta}{\delta
  r_{j}^{(p)}(\vec{x}',t)}V\left({\bf r}^{(p)}(\vec{x}',t)-{\bf
R}(\vec{x},t)\right)\Bigg\}\label{Xi}
\end{eqnarray}
where the summation over repeated Cartesian component indices is implied and the matrix manifolds action is defined 
\begin{eqnarray}
A_{2}\left[{\bf r}^{(p)}({\vec x},t),{\hat {\bf r}}^{(p)}({\vec x},t)\right]=&-&\sum_{p=1}^{M}\int
d^{D}x\int dt\Bigg\{i{\hat r}^{(p)}_{j}({\vec
  x},t)\left[\xi_{0}\frac{\partial}{\partial t}r^{(p)}_{j}({\vec
  x},t)-\epsilon\Delta_{x}r^{(p)}_{j}({\vec
  x},t)\right]\nonumber\\
&-&\sum_{m=1}^{M}\int d^{D}x' i{\hat r}^{(p)}_{j}({\vec
  x},t)\frac{\delta}{\delta r^{(p)}_{j}({\vec
  x},t)}V\left[{\bf r}^{(p)}(\vec{x},t)-{\bf r}^{(m)}(\vec{x}',t)\right]\nonumber\\&+&T\xi_{0}\left[i{\hat r}^{(p)}_{j}({\vec
  x},t)\right]^{2}\Bigg\}\label{matrixaction}.
\end{eqnarray}
The representation (\ref{GF1})-(\ref{matrixaction}) is very useful for
performing transformations to collective variables as well as integration over
a subset of variables. In our case we make the transformation to the matrix
density $\rho({\bf r},t)$ and the matrix response field density $\pi({\bf
  r},t)$:
\begin{eqnarray}
\rho({\bf r},t)=\sum_{p=1}^{M}\int d^{D}x\:\delta\left({\bf r}-{\bf
    r}^{(p)}({\vec x},t)\right)\label{rho}
\end{eqnarray}
\begin{eqnarray}
\pi({\bf
  r},t)=\sum_{p=1}^{M}\sum_{j=1}^{d}\int d^{D}x\:i{\hat r}^{(p)}_{j}({\vec
  x},t)\nabla_{j}\delta\left({\bf r}-{\bf r}^{(p)}({\vec
    x},t)\right)\label{pi}.
\end{eqnarray} 
The transformation to the collective variables has already been used before
for the dynamics of semi-dilute polymer solutions \cite{23} and melts
\cite{6,24}. The principal aim now is to integrate in the influence functional
(\ref{Xi}) over the collective variables (\ref{rho})-(\ref{pi}), and as a
result to find the effective dynamic of the test chain/manifold in the matrix.

The representation of the influence functional in terms of the collective
fields $\rho$ and $\pi$
has the form
\begin{eqnarray}
\Xi\left[{\bf R},{\hat {\bf R}}\right]=&-&\ln \int D\rho({\bf k},t)D\pi({\bf
  k},t)\nonumber\\
&\times&\exp\Bigg\{W[\rho,\pi]-\int dt\int\frac{d^{d}k}{(2\pi)^{d}}\pi (-{\bf
  k},t)\rho({\bf k},t)V({\bf k})\nonumber\\
&+&\int d^{D}x\int dt\:i{\hat R}_{j}({\vec
  x},t)\int\frac{d^{d}k}{(2\pi)^{d}}ik_{j}V({\bf k})\rho({\bf k},t)\exp\{i{\bf
  k}{\bf R}({\vec x},t)\}\nonumber\\
&-&\int d^{D}x\int dt\int\frac{d^{d}k}{(2\pi)^{d}}\pi ({\bf k},t)V(-{\bf
  k})\exp\{i{\bf k}{\bf R}({\vec x},t)\}\Bigg\}\label{Xirho}
\end{eqnarray} 
with a potential $W$ that depends only on properties of the free system 
\begin{eqnarray}
W\{\rho,\pi \}&=&\ln \int\prod_{p=1}^{M}D{\bf r}^{(p)}D{\hat {\bf
    r}}{(p)}\exp\left\{A_{0}\left[{\bf r}^{(p)},{\hat {\bf
        r}}^{(p)}\right]\right\}\nonumber\\
&\times&\delta\left(\rho({\bf r},t)-\sum_{p=1}^{M}\int d^{D}x\:\delta\left({\bf r}-{\bf
    r}^{(p)}({\vec x},t)\right)\right)\nonumber\\
&\times&\delta\left(\pi ({\bf
  r},t)-\sum_{p=1}^{M}\sum_{j=1}^{d}\int d^{D}x\:i{\hat r}^{(p)}_{j}({\vec
  x},t)\nabla_{j}\delta\left({\bf r}-{\bf r}^{(p)}({\vec
    x},t)\right)\right)\label{W}
\end{eqnarray}
and $A_{0}$ the free system action
\begin{eqnarray}
A_{0}\left[{\bf r}^{(p)},{\hat {\bf r}}^{(p)}\right]=\sum_{p=1}^{M}\int
d^{D}xdt\left\{i{\hat r}_{j}^{(p)}\left[\xi_{0}\frac{\partial}{\partial
      t}r_{j}^{(p)}-\epsilon\Delta_{x}r_{j}^{(p)}\right]+T\xi_{0}\left[i{\hat
      {\bf r}}^{(p)}\right]^{2}\right\}\label{212}.
\end{eqnarray}
The Eqs.~(\ref{W})-(\ref{212}) can be brought into a very compact form in terms of the 2 -
dimensional column variables $\rho_{\alpha}$ and $\chi_{\alpha}$
\begin{eqnarray}
\rho_{\alpha}({\bf k},t)=\begin{pmatrix}
  \rho({\bf k},t) \\ \pi({\bf k},t)
\end{pmatrix}\label{vec}
\end{eqnarray}
and 
\begin{eqnarray}
\chi_{\alpha}({\bf k},t)=\begin{pmatrix}
-\int d^{D}x\:i{\hat R}_{j}({\vec x},t)ik_{j}V(-{\bf k})\exp\left[-i{\bf
    k}{\bf R}({\vec x},t)\right] \\
-\int d^{D}x\:V({\bf k})\exp\left[-i{\bf k}{\bf R}({\vec x},t)\right]
\end{pmatrix}\label{chi}
\end{eqnarray} 
where $\alpha=0,1$.

By making use of eqs.~(\ref{vec})-(\ref{chi}) in eq.~(\ref{Xirho}) we arrive
at
\begin{multline}
\Xi\left[{\bf R},{\hat {\bf R}}\right]=-\ln
\int\prod_{\alpha}D\rho_{\alpha}({\bf
  k},t)\exp\Bigg\{W\left\{\rho_{\alpha}\right\}\\
-\frac{1}{2}\int dt\int
\frac{d^{d}k}{(2\pi)^{d}}\rho_{\alpha}({\bf k},t)U_{\alpha\beta}({\bf
  k})\rho_{\beta}(-{\bf k},t)\\+\int dt\int
\frac{d^{d}k}{(2\pi)^{d}}\chi_{\alpha}(-{\bf k},t)\rho_{\alpha}({\bf
  k},t)\Bigg\}\label{Xicomp}
\end{multline}
with the $2\times 2$ - interaction matrix 
\begin{eqnarray}
U_{\alpha\beta}=\begin{pmatrix}
0 & V({\bf k}) \\
V({\bf k}) & 0
\end{pmatrix}.
\end{eqnarray}
In eq.~(\ref{Xicomp}) the summation over repeated Greek indices is implied.

Up to now all calculations have been exact. In order to proceed we should specify
the "potential" $W\left\{\rho_{\alpha}\right\}$. Unfortunately the exact form
for $W$ is not known explicitly, but in ref.~\cite{25} the systematic
expansion in the $\rho_{\alpha}$ - variables was given for the first
time. Here we will use this expansion only up to the second order, which
corresponds to the dynamical RPA
\begin{eqnarray}
W\left\{\rho_{\alpha}\right\}=W\left\{\left<\rho_{\alpha}\right>_{0}\right\}+\frac{1}{2!}\int
dt dt'\frac{d^{d}k}{(2\pi)^{d}}\rho_{\alpha}({\bf k},t)W_{\alpha\beta}^{(2)}({\bf
  k};t-t')\rho_{\beta}(-{\bf k},t')+\ldots\label{expa}
\end{eqnarray}
where
\begin{eqnarray}
W_{\alpha\beta}^{(2)}({\bf k},t)=-\left(F^{-1}\right)_{\alpha\beta}({\bf
  k},t)\label{2p}
\end{eqnarray}
and $\left(F^{-1}\right)_{\alpha\beta}({\bf
  k},t)$ stands for inversion of the matrix of the free system
\begin{eqnarray}
F_{\alpha\beta}\left({\bf k},t\right)=\begin{pmatrix}
F_{00}({\bf k},t) & F_{01}({\bf k},t) \\
F_{10}({\bf k},t) & 0
\end{pmatrix}\label{219}.
\end{eqnarray}
In the $2\times 2$ - matrix (\ref{219}) $F_{00}({\bf k},t)$, $F_{01}({\bf
  k},t)$ and $F_{10}({\bf k},t)$ are time correlation, retarded response and
advanced response functions correspondingly.

The Gaussian approximation in eq.~(\ref{expa}) corresponds to the dynamical
random phase approximation (RPA) \cite{25}. The RPA makes the integration over
$\rho_{\alpha}$ in eq.~(\ref{Xicomp}) analytically amenable and as a result
for the GF we have
\begin{eqnarray}
Z\{\ldots\}=\int D{\bf R}({\vec x},t)D{\hat {\bf R}}({\vec
  x},t)\exp\left\{-A\left[{\bf R},{\hat {\bf R}}\right]\right\}\label{220}
\end{eqnarray}
where
\begin{multline}
A\left[{\bf R},{\hat {\bf R}}\right]=A_{0}\left[{\bf R},{\hat {\bf
        R}}\right]\\
-\frac{1}{2}\int d1 d1'\:i{\hat
  R}_{j}(1)\int\frac{d^{d}k}{(2\pi)^{d}}k_{j}k_{p}|V({\bf
  k})|^{2}\exp\left\{i{\bf k}\left[{\bf R}(1)-{\bf R}(1')\right]\right\}i{\hat
  R}_{p}(1')S_{00}({\bf k},t-t')\\
+\int d1 d1'\:i{\hat
  R}_{j}(1)\int\frac{d^{d}k}{(2\pi)^{d}}ik_{j}|V({\bf
  k})|^{2}\exp\left\{i{\bf k}\left[{\bf R}(1)-{\bf
      R}(1')\right]\right\}S_{01}({\bf k},t-t')\\
-\int d1 d1'\:i{\hat
  R}_{j}(1)\int\frac{d^{d}k}{(2\pi)^{d}}ik_{j}V({\bf k})\exp\left\{i{\bf k}\left[{\bf R}(1)-{\bf
      R}(1')\right]\right\}\delta(t-t')\label{A}
\end{multline}
where we used a short hand notation in the form $1\equiv ({\vec x},t)$, and
$A_{0}$ is the action of the free test manifold
\begin{eqnarray}
A_{0}\left[{\bf R},{\hat {\bf R}}\right]=-\int d1\left\{i{\hat
  R}_{j}(1)\left[\xi_{0}\frac{\partial}{\partial
    t}R_{j}(1)-\Delta_{x}R_{j}(1)\right]+T\xi_{0}\left[i{\hat
    R}_{j}(1)\right]^{2}\right\}\label{222}.
\end{eqnarray}
$S_{00}({\bf k},t)$ and $S_{01}({\bf k},t)$ are the corresponding elements
of the dynamic matrix in RPA, which is given by
\begin{eqnarray}
S_{\alpha\beta}({\bf
  k},t)=\left[U+\left(F^{-1}\right)\right]_{\alpha\beta}^{-1}({\bf
  k},t)\label{223}.
\end{eqnarray}
The GF (\ref{220}) determines the dynamics of the self-avoiding test manifold,
which is modulated by the melt fluctuations given in the dynamical
RPA.\footnote{In ref.~\cite{6} the self-avoidance was dropped and instead
  of RPA-correlators the full correlators were introduced.}

To go beyond RPA anharmonicities in the expansion
(\ref{expa})  must be taken into account and follow the renormalized perturbation theory which was worked
out in ref.~\cite{25} for the glass transition problem. In case the
temperature is much higher than the glass transition temperature the
fluctuations in a polymer melt are described by the RPA reasonably well
\cite{26}.
\subsection{The selfconsistent Hartree approximation}
The resulting action (\ref{A}) includes the test manifold variables in a
highly non-linear way. In order to handle this we use the Hartree-type
approximation [1,8,9]. In the selfconsistent Hartree approximation the real
MSR - action is replaced by a Gaussian action in such a way that all terms
which include more than two fields $R_{j}({\vec x},t)$ or/and ${\hat R}_{j}({\vec
  x},t)$ are written in all possible ways as products of pairs of $R_{j}({\vec
  x},t)$ or ${\hat R}_{j}({\vec
  x},t)$, coupled to selfconsistent averages of the remaining fields. As a
result the Hartree - action is a Gaussian functional with coefficients, which
could be represented in terms of correlation and response functions. The
calculations are straightforward and the details can be found in the Appendix
B of the ref.~[8]. The only difference is that here we deal with the
self-avoiding $D$ - dimensional manifold (see the last term in eq.~(21)) in
the $d$ - dimensional space and the collective dynamics of the melt is treated
in the framework of the dynamical RPA. The resulting GF takes the form
\begin{multline}
Z\{\cdots \}=\int D{\bf R}D{\hat {\bf R}}\exp\Big\{-A_{0}[{\bf R},{\hat {\bf
    R}}]\\
+\int d^{D}xd^{D}x'\int_{-\infty}^{\infty}dt\int_{-\infty}^{t}dt'\:i{\hat
  R}_{j}({\vec x},t)R_{j}({\vec x}',t')\lambda({\vec x},{\vec x}';t,t')\\
-\int d^{D}xd^{D}x'\int_{-\infty}^{\infty}dt\int_{-\infty}^{t}dt'\:i{\hat
  R}_{j}({\vec x},t)R_{j}({\vec x},t)\lambda({\vec x},{\vec x}';t,t')\\
+\frac{1}{2}\int d^{D}xd^{D}x'\int_{-\infty}^{\infty}dt\int_{-\infty}^{\infty}dt'\:i{\hat
  R}_{j}({\vec x},t)i{\hat R}_{j}({\vec x}',t')\chi({\vec x},{\vec
  x}';t,t')\Big\}\label{226}
\end{multline}
where 
\begin{eqnarray}
\lambda({\vec x},{\vec x}';t,t')&=&\frac{1}{d}G({\vec x},{\vec
  x}';t,t')\int\frac{d^{d}k}{(2\pi)^{d}}k^{4}|V({\bf k})|^{2}F({\bf k};{\vec
  x},{\vec x}';t,t')S_{00}({\bf k};t,t')\nonumber\\ 
&-&\int\frac{d^{d}k}{(2\pi)^{d}}k^{2}\left[|V({\bf k})|^{2}S_{01}({\bf
    k};t,t')-V({\bf k})\delta(t-t')\right]F({\bf k};{\vec
  x},{\vec x}';t,t')\label{227}
\end{eqnarray}
and
\begin{eqnarray}
\chi({\vec x},{\vec x}';t,t')=\int\frac{d^{d}k}{(2\pi)^{d}}k^{2}|V({\bf k})|^{2}F({\bf k};{\vec
  x},{\vec x}';t,t')S_{00}({\bf k};t,t')\label{228}.
\end{eqnarray}
In eqs.~(\ref{227})-(\ref{228}) the response function 
\begin{eqnarray}
G({\vec x},{\vec
  x}';t,t')=\left<i{\hat {\bf R}}({\vec x}',t'){\bf R}({\vec x},t)\right>
\end{eqnarray}
and the density correlator
\begin{eqnarray}
F({\bf k};{\vec
  x},{\vec x}';t,t')=\exp\left\{-\frac{k^{2}}{d}Q({\vec
  x},{\vec x}';t,t')\right\}\label{230}
\end{eqnarray}
with 
\begin{eqnarray}
Q({\vec
  x},{\vec x}';t,t')\equiv\left<{\bf R}({\vec x},t){\bf R}({\vec
    x},t)\right>-\left<{\bf R}({\vec x},t){\bf R}({\vec
    x}',t')\right>\label{231}
\end{eqnarray}
are specified. The pointed brackets denote the selfconsistent averaging with
the Hartree-type GF (26).

It is obvious that for the case under consideration the time homogeneity and
the Fluctuation-Dissipation Theorem (FDT) hold, then
\begin{eqnarray}
G({\vec x},{\vec
  x}';t-t')=T^{-1}\frac{\partial}{\partial t'}Q({\vec
  x},{\vec x}';t-t')\:\:\:\:t>t'\label{232}
\end{eqnarray}
\begin{eqnarray}
S_{01}({\bf k};t-t')=T^{-1}\frac{\partial}{\partial t'}S_{00}({\bf k};t-t')\:\:\:\:t>t'\label{233}.
\end{eqnarray}
By making use of eqs.~(\ref{232}) and (\ref{233}) in
eqs.~(\ref{226})-(\ref{231}) and after integration by parts with respect to
the time argument $t'$, we obtain 
\begin{eqnarray}
Z\left\{\cdots\right\}&=&\int DR_j({\vec x},t)D{\hat
  R}_j({\vec x},t)\nonumber\\
&\times&\exp\Bigg\{\int_{0}^{N}d^{D}x\:d^{D}x'\int_{-\infty}^{\infty} dtdt'\:i{\hat
    R}_{j}({\vec x},t)\Bigg[\xi_{0}\delta(t-t')\delta({\vec x}-{\vec x}')+\nonumber\\
&\:&+\:\Theta(t-t')\frac{1}{T}\int\frac{d^{d}k}{(2\pi)^{d}}k^{2}|V({\bf k})|^{2}F({\bf
    k};{\vec x},{\vec x}';t-t')S_{00}({\bf k};t-t')\Bigg]\frac{\partial}{\partial
  t'}R_{j}({\vec x}',t')\nonumber\\
&-&\int_{0}^{N}d^{D}x\:d^{D}x'\int_{-\infty}^{\infty}dt\:i{\hat
  R}_{j}({\vec x},t)\Bigg[\varepsilon\delta({\vec x}-{\vec
  x}')\Delta_{x}+\int\frac{d^{d}k}{(2\pi)^{d}}k^{2}\left[|V({\bf k})|^{2}\frac{1}{T}S_{\rm st}({\bf
k})-V({\bf k})\right]\nonumber\\
&\:&\:\times\left[F_{\rm st}({\bf
    k};{\vec x},{\vec x}')-\delta({\vec x}-{\vec x}')\int_{0}^{N}d^{D}x^{''}F_{\rm st}({\bf
k};{\vec x},{\vec x}^{''})\right]\Bigg]R_{j}({\vec x}',t)\nonumber\\&+&T\int_{0}^{N}d^{D}x\:d^{D}x'\int_{-\infty}^{\infty}
dtdt'\Bigg[\xi_{0}\delta(t-t')\delta({\vec x}-{\vec x}')+\Theta(t-t')\nonumber\\&\:&\:\frac{1}{T}\int\frac{d^{d}k}{(2\pi)^{d}}k^{2}|V({\bf k})|^{2}F({\bf
    k};{\vec x},{\vec x}';t-t')S_{00}({\bf k};t-t')\Bigg]i{\hat R}_{j}({\vec x},t)i{\hat
R}_{j}({\vec x}',t')\label{234}
\end{eqnarray}
where $F_{\rm st}({\bf k};{\vec x},{\vec x}')$ is the static density correlation
function.

The generalized Rouse equation (GRE), which directly follows from the GF
(\ref{234}), has the form
\begin{multline}
\xi_{0}\frac{\partial}{\partial t}R_{j}({\vec x},t)+\int
d^{D}x'\int_{0}^{t}dt'\Gamma({\vec x},{\vec x}';t-t')\frac{\partial}{\partial
  t'}R_{j}({\vec x}',t')\\
-\int d^{D}x'\Omega({\vec x},{\vec x}')R_{j}({\vec x}',t)={\cal F}({\vec
  x},t)\label{235}
\end{multline}
with the memory function
\begin{eqnarray}
\Gamma({\vec x},{\vec
  x}';t)=\frac{1}{T}\int\frac{d^{d}k}{(2\pi)^{d}}k^{2}|V({\bf k})|^{2}F({\bf k};{\vec
  x},{\vec x}';t)S_{00}({\bf k},t)\label{236}
\end{eqnarray} 
and the effective static elastic susceptibility  
\begin{multline}
\Omega({\vec x ,{\vec x}'})=\epsilon\Delta({\vec x}-{\vec
  x}')\delta_{x}\\-\int\frac{d^{d}k}{(2\pi)^{d}}k^{2}{\cal V}({\bf
  k})\left[F_{\rm st}({\bf k};{\vec x}'{\vec x}')-\delta({\vec x}-{\vec x}')\int
  d^{D}x''F_{\rm st}({\bf k};{\vec x},{\vec x}'')\right]\label{237}
\end{multline}
and the random force ${\cal F}({\vec x},t)$ has the correlator 
\begin{eqnarray}
\left<{\cal F}({\vec x},t){\cal F}({\vec
    x}',t')\right>=2T\delta_{ij}\left[\xi_{0}\delta({\vec x}-{\vec
    x}')\delta(t-t')+\Theta(t-t')\Gamma({\vec x},{\vec
    x}';t-t')\right]\label{238}.
\end{eqnarray}
In eq.~(\ref{237}) the effective interaction function
\begin{eqnarray}
{\cal V}({\bf k})=V({\bf k})\left[1-\frac{1}{T}V({\bf k})S_{\rm st}({\bf
    k})\right]\label{239}
\end{eqnarray}
gains the standard screened form \cite{14}
\begin{eqnarray}
{\cal V}({\bf k})=\frac{V({\bf k})}{1+V({\bf k})F_{\rm st}^{(0)}({\bf k})/T}\label{240}
\end{eqnarray}
(where $F_{\rm st}^{0}({\bf k})$ is the free system correlator), if the standard RPA-result is used for the melts static correlator
\begin{eqnarray}
S_{\rm st}({\bf k})=\frac{F_{\rm st}^{(0)}({\bf k})}{1+V({\bf k})F_{\rm st}^{(0)}({\bf
    k})/T}\label{241}.
\end{eqnarray} 
The GRE (\ref{235})-(\ref{238}) is the generalization of the corresponding
equations given in \cite{6} for the case of a test manifold with self-excluded
volume. On the other hand the collective dynamics of the melt is treated in
the framework of the RPA. This is a good starting point for a simultaneous
consideration of the static and dynamic behavior of the test manifold. One
should expect, for example, that the reactive and dissipative forces in
eq.~(\ref{235}) are screened out in different ways. As we will show in the next
section this is really the case. Explicitly stated, the test manifold could be statically
Gaussian, but dynamically it could follow a renormalized Rouse dynamics.
\section{Static and dynamic behavior of the test manifold}
For the following discussion, it is convenient to perform a
Fourier-transformation with respect to the variable ${\vec x}$. We define
e.g. the Rouse-mode time correlation function
\begin{eqnarray}
C({\vec p},t)=\frac{1}{{\cal N}}\int d^{D}x C({\vec
  x},t)e^{-i\frac{2\pi}{N}{\vec x}\cdot{\vec p}}\label{31}
\end{eqnarray}
and its inverse transformation
\begin{eqnarray}
C({\vec
  x},t)=\int d^{D}p C({\vec p},t)e^{i\frac{2\pi}{N}{\vec x}\cdot{\vec p}}\label{32}
\end{eqnarray}
where ${\cal N}=N^{D}$ is the total number of "monomers" (or beads).

Then eq.~(\ref{235}) leads to the result
\begin{eqnarray}
\xi_{0}\frac{\partial}{\partial t}C({\vec p},t)+{\cal
  N}\int_{0}^{t}dt'\Gamma({\vec p},t-t')\frac{\partial}{\partial t'}C({\vec
  p},t')+\Omega({\vec p})C({\vec p},t)=0\label{33}
\end{eqnarray}
where $\Gamma({\vec p},t)$ and $\Omega({\vec p})$ are the Fourier transformations
of the memory function (\ref{236}) and the susceptibility (\ref{237})
respectively. Below we shall analyse both of them explicitly.
\subsection{Static properties}
The static limit $t\rightarrow 0^{+}$ of eq.~(\ref{33}) can be implemented, if
we take into account 
\begin{eqnarray}
\xi_{0}\left[\frac{\partial}{\partial t}C({\vec x},t)\right]_{t\rightarrow
  0^{+}}=T\xi_{0}G({\vec x},t\rightarrow 0^{+})=-dT\delta({\vec x})\label{34}
\end{eqnarray}
where the FDT (\ref{232}) as well as the initial condition for the response
function (see eq.~(31) in \cite{6} or eq.~(3.12) in \cite{1}) have been
used. Then the formal solution for the static Rouse mode correlator, $C({\vec
  p})=\left<{\bf R}({\vec p}){\bf R}(-{\vec p})\right>$, becomes
\begin{eqnarray}
C({\vec p})=\frac{d}{{\cal N}\left[\frac{d}{l^{2}}\left(\frac{2\pi{\vec
          p}}{N}\right)^{2}+\Sigma({\vec p})\right]}\label{35}
\end{eqnarray}
The eq.~(\ref{35}) has the Dyson-like form where the "self-energy" is given by
\begin{eqnarray}
\Sigma({\vec p})={\cal N} \int\frac{d^{d}k}{(2\pi)^{d}}k^{2}\frac{V({\bf
    k})/T}{1+V({\bf k})F_{\rm st}^{(0)}({\bf k})/T}\left[F_{\rm st}({\bf k},{\vec
    p})-F_{\rm st}({\bf k},{\vec p}=0)\right]\label{36}.
\end{eqnarray}
The static correlator $C({\bf p})$ is parametrized by the wandering exponent
$\zeta$ (see eq.~(\ref{231}))
\begin{eqnarray}
Q_{\rm st}({\vec x})=\int d^{D}p\left[1-e^{-i\frac{2\pi}{N}{\vec p}\cdot{\vec
      x}}\right]C({\vec p})\propto x^{2\zeta}\label{37}.
\end{eqnarray}
In its turn the test manifolds static correlator in eq.~(\ref{36}) is 
\begin{eqnarray} 
F_{\rm st}({\bf k};{\vec p})=\frac{1}{{\cal N}}\int
d^{D}x\exp\left\{-\frac{k^{2}l^{2}}{2d}x^{2\zeta}-i\frac{2\pi}{N}{\vec
      x}\cdot{\vec p}\right\}\label{38}.
\end{eqnarray}
As a good approximation for the free system correlator, $F_{\rm st}^{(0)}({\bf
  k})$, we will use the Pad\'e formula
\begin{eqnarray}
F_{\rm st}^{(0)}({\bf k})=\frac{\rho{\cal N}}{1+(kl)^{d_{f}^{o}}{\cal
    N}\gamma_{1}(d,D)}\label{39}
\end{eqnarray}
where $\rho$ is the averaged bead density and 
\begin{eqnarray}
\gamma_{1}(d,D)=\left[\frac{S_{D}}{2\zeta_{0}}(2d)^{d_{f}^{o}/2}\Gamma\left(\frac{d_{f}^{o}}{2}\right)\right]^{-1}\label{310}
\end{eqnarray}
with the Gaussian fractal dimension $d_{f}^{o}=2D/(2-D)$, the Gaussian
exponent $\zeta_{0}=(2-D)/2$, $S_{D}=2\pi^{D/2}/\Gamma(D/2)$ the volume of the
unit spere and the
gamma-function $\Gamma(x)$. The system of eqs.~(\ref{35})-(\ref{310}) can be
analysed self-consistently.

Let us start from the calculation of the "self-energy" (\ref{36}). Because
$\beta VF_{\rm st}^{(0)}\gg 1$, the effective screened interaction is proportional
to $1/F_{\rm st}^{(0)}({\bf k})$. Then using eqs.~(\ref{38})-(\ref{310}) in
eq.~(\ref{36}) and performing the integration over ${\bf k}$ first yields
\begin{eqnarray}
\Sigma(p)=\Sigma_{1}(p)+\Sigma_{2}(p)\label{311}
\end{eqnarray}
with
\begin{eqnarray}
\Sigma_{1}(p)&=&-c_{1}\left(\frac{\pi p}{N}\right)^{\zeta(d+2)}\nonumber\\
\Sigma_{2}(p)&=&-c_{2}\left(\frac{\pi p}{N}\right)^{\zeta(d+d_{f}^{o}+2)-D}\label{312}
\end{eqnarray}
where $c_{1}$ and $c_{2}$ are given by
\begin{multline}
c_{1}=\frac{S_{d}S_{D}}{l^{2}(2\pi)^{d}}\cdot\frac{(2d)^{\frac{d+2}{2}}}{2\left(\rho
    l^{d}\right)}\Gamma\left(\frac{d}{2}+1\right)\\
\times\frac{1}{(\pi
  p)^{D}}\int_{0}^{\pi
  p}dt\:t^{D-1-\zeta(d+2)}\left[1-\Gamma\left(\frac{D}{2}\right)\left(\frac{1}{t}\right)^{\frac{D-2}{2}}J_{\frac{D-2}{2}}(2t)\right]\label{313} 
\end{multline}
and
\begin{multline}
c_{2}=\frac{S_{d}S_{D}}{l^{2}(2\pi)^{d}}\cdot\frac{(2d)^{\frac{d+d_{f}^{o}}{2}+1}}{2\left(\rho
    l^{d}\right)}\gamma_{1}\cdot\Gamma\left(\frac{d+d_{f}^{o}}{2}+1\right)\\
\times\int_{0}^{\pi
  p}dt\:t^{D-1-\zeta(d+d_{f}^{o}+2)}\left[1-\Gamma\left(\frac{D}{2}\right)\left(\frac{1}{t}\right)^{\frac{D-2}{2}}J_{\frac{D-2}{2}}(2t)\right]\label{314}. 
\end{multline}
Here $J_{D/2-1}(x)$ is the Bessel function.

We assume that $p={\cal O}(1)$ but $(p/N)\ll 1$. Physically, the condition for
the exponent $\zeta$ comes from the balance between the entropic and the
interaction terms in the denominator of eq.~(\ref{35}). But one should also
be wary about the self-consistency condition (\ref{37}) otherwise the result
could be different.
\renewcommand{\theenumi}{\roman{enumi}}
\begin{enumerate}
\item Let us, e.g. assume that 
\begin{eqnarray}
\zeta(d+2)<\zeta(d+d_{f}^{o}+2)-D\label{315}.
\end{eqnarray}
In this case $\Sigma_{1}$ can compensate the entropic term proportional to $(\pi
p/N)^{2}$ in eq.~(\ref{35}) and one should claim $\zeta(d+2)=2$ or
\begin{eqnarray}
\zeta=\frac{2}{d+2}
\label{316}.
\end{eqnarray}
Then condition (\ref{315}) yields
\begin{eqnarray}
d<\frac{2D}{2-D}\equiv d_{\rm uc}\label{317}
\end{eqnarray}
where $d_{\rm uc}$ is the upper critical dimension in a melt \cite{19}. The
results (\ref{316})-(\ref{317}) was obtained first in \cite{19}.

Nevertheless, one can see that the exponent (\ref{316}) does not fulfill the
self-consistency condition (\ref{37})
\begin{eqnarray}
Q_{{\rm st}}(x)\propto\int d^{D}p\left[1-\exp\left(-i\frac{2\pi}{N}{\vec p}{\vec
      x}\right)\right]{\Big /}\left(\frac{\pi p}{N}\right)^{2}\label{317b}.
\end{eqnarray}
We will come to the contravention of the condition (\ref{37}) also in the case
where $\zeta(d+2)>\zeta(d+d_{f}^{o}+2)-D$.
\item The only way to satisfy the eqs.~(\ref{35})-(\ref{310}) is to impose
  on the exponents in eq.~(\ref{312}) the condition
\begin{eqnarray}
\zeta(d+2)=\zeta(d+d_{f}^{o}+2)-D>2\label{318}.
\end{eqnarray}
In this case $\zeta d_{f}^{o}-D=0$ or 
\begin{eqnarray}
\zeta=\zeta_{0}=\frac{2-D}{2}\label{319}.
\end{eqnarray}
On the other hand eq.~(\ref{318}) yields
\begin{eqnarray}
d>\frac{2D}{2-D}=d_{\rm uc}\label{320}
\end{eqnarray}
i.e. the Gaussian solution (\ref{319}) is self-consistent at $d>d_{\rm uc}$. In
this case the entropic term dominates : $(\pi p/N)^{2}\gg |\Sigma_{1}|\propto
|\Sigma_{2}|$. The criterion (\ref{320}) is equivalent to the embedding
condition $d_{f}^{o}<d$ and can be represented in the form
\begin{eqnarray}
D<D_{s}=\frac{2d}{2+d}\label{323}.
\end{eqnarray}
The spectral critical dimension $D_{s}$ was discussed first in \cite{19}.
\item At $d=d_{\rm uc}$
\begin{eqnarray}
\zeta_{0}(d+2)=\zeta_{0}(d+d_{f}^{o}+2)-D=2\label{321}
\end{eqnarray}
and all terms have the same order : $(\pi p/N)^{2}\propto |\Sigma_{1}|\propto
|\Sigma_{2}|$. The system is {\it marginally stable} and the stability condition
is given by
\begin{eqnarray}
\left[\frac{4d_{\rm uc}}{l^{2}}-c_{1}(d_{\rm uc},D)-c_{2}(d_{\rm uc}D)\right]>0\label{322}
\end{eqnarray}
which is always valid if $\rho l^{d}\gg 1$.
\item If $\zeta_{0}(d+2)=\zeta(d+d_{f}^{o}+2)-D<2$, i.e. $d<d_{\rm uc}$, the tems
  $\Sigma_{1}$ and $\Sigma_{2}$ overwhelm the entropic one, $(\pi
  p/N)^{2}\ll|\Sigma_{1}|\propto |\Sigma_{2}|$, and the system becomes
  unstable. Hence, at $d<d_{\rm uc}$ the manifold is saturated in the melt,
  i.e. it loses its fractal nature and becomes compact \cite{18}. 
\end{enumerate} 
\subsection{Dynamic properties at $d\ge d_{\rm uc}$}
We consider now the dynamics at $d\ge d_{\rm uc}$. There are two dynamic
exponents, $z$ and $w$. The exponent $z$ measures the time dependence of a
monomer displacement, i.e. 
\begin{eqnarray}
Q(t)=\int d^{D}p\int_{a-i\infty}^{a+i\infty}\frac{ds}{2\pi
  i}\left[1-e^{st}\right]C({\vec p},s)\propto t^{2z}\label{324}
\end{eqnarray}
and the exponent $w$ measures the same for the center of mass
\begin{eqnarray}
Q_{\rm cm}(t)=\lim_{p\rightarrow 0}\int_{a-i\infty}^{a+i\infty}\frac{ds}{2\pi
  i}\left[1-e^{st}\right]C({\vec p},s)\propto t^{w}\label{325}.
\end{eqnarray} 
In eqs.~(\ref{324})-(\ref{325}) $C({\vec p},s)=\left<|{\bf R}({\vec
    p},s)|^{2}\right>$ is the Rouse-Laplace component of the correlator
$C({\vec x},t)$ and the integral over $s$ is taken along a straight line with
Re$(s)=a$, so that the function is analytic at Re$(s)>a$.

The formal solution of eq.~(\ref{33}) for $C({\vec p},s)$ is given by 
\begin{eqnarray}
\newcommand{\D}{\displaystyle}
\normalsize
C({\vec p},s)=\frac{C_{\rm st}({\vec p})}{s+{\D \frac{\epsilon\left(\frac{2\pi
        p}{N}\right)^{2}}{\xi_{0}+{\cal N}\Gamma({\vec p},s)}}}\label{326}
\end{eqnarray}
where we have taken into account that at $d>d_{\rm uc}$ the manifold is Gaussian,
i.e. $\Omega(p)=\epsilon (2\pi p)/N)^{2}$ and $C_{\rm st}(p)=l^{2}(N/2\pi
p)^{2}/{\cal N}$. In eq.~(\ref{326}) $\Gamma({\vec p},s)$ is the Rouse-Laplace component of the memory function (\ref{236}).

In order to calculate $\Gamma({\vec p},s)$ one needs analytically tractable
approximations for the matrix density correlator $S_{00}({\bf k},t)$ in RPA
and for the test manifold correlator $F({\bf k};{\vec x},t)$. At
$(kl)^{d_{f}^{o}}{\cal N}\ll 1$ we can use for the free system density
correlator \cite {14}
\begin{eqnarray}
F_{00}^{(0)}({\bf k},t)=F_{\rm st}^{(0)}\exp\left[-k^{2}D(k)t]\right]\label{327}
\end{eqnarray}
where $D(k)=D_{0}/g(k)$, $D_{0}=T/\xi_{0}$, $g(k)$ is the Gaussian manifold
static structure factor and $F_{\rm st}^{(0)}$ is the corresponding static
correlator. Using eq.~(\ref{327}) in the RPA-dynamic matrix (\ref{223}),
yields for the matrix density correlator 
\begin{eqnarray}
S_{00}({\bf k},t)=S_{\rm st}(k)\exp\left\{-k^{2}D_{\rm coop}(k)t\right\}\label{328}
\end{eqnarray}
where $S_{\rm st}(k)$ is the static RPA-correlator,
$D_{\rm coop}(k)=\rho/\xi_{o}\chi_{\rm st}(k)$ and $\chi_{\rm st}^{-1}(k)=T\left[F_{\rm st}^{(0)}(k)\right]^{-1}+V({\bf k})$. At $(kl)^{d_{f}^{o}}{\cal N}\ll 1$ we
have $F_{\rm st}^{(0)}\approx \rho{\cal N}$ and $\chi_{\rm st}^{-1}\approx V({\bf k})$. Hence
\begin{eqnarray}
D_{\rm coop}(k)\approx\frac{\rho}{\xi_{0}}V({\bf k})\label{329}.
\end{eqnarray}
At $(kl)^{d_{f}^{(0)}}{\cal N}\gg 1$ we probe only the local motion and, as a
result, the dynamics is mainly determined by the single manifold behavior. So,
the RPA-correlator $S_{00}({\bf k},t)$ is approximated by
\begin{eqnarray}
S_{00}({\bf k},t)=S_{\rm st}({\bf k})\times\left\{\begin{array}{r@{\quad,\quad}l}
\exp\left\{-k^{2}D_{\rm coop}({\bf k})t\right\}\:\: & (kl)^{d_{f}^{0}}{\cal  
N}\ll 1\\
\exp\left\{-\frac{k^{2}l^{2}}{2d}\left(\frac{t}{\tau_{0}}\right)^{2z_{0}}\right\}  
& (kl)^{d_{f}^{0}}{\cal N}\gg 1\end{array}\right.\label{330}
\end{eqnarray}
where $\tau_{0}=\xi_{0}l^{2}/Td$ and $z_{0}=(2-D)/4$ is the Gaussian
z-exponent. 

The corresponding Ansatz for the test manifold yields
\begin{eqnarray}
F({\bf k};{\vec x};t)=F_{\rm st}({\bf k},{\vec
  x})\times\left\{\begin{array}{r@{\quad,\quad}l}\exp\left\{-k^{2}D_{G}t\right\}\quad\quad\:\:  
& (kl)^{d_{f}^{0}}{\cal N}\ll 1\\
\exp\left\{-\frac{k^{2}l^{2}}{2d}\left(\frac{t}{\tau_{0}}\right)^{2z}\right\}
& (kl)^{d_{f}^{0}}{\cal N}\gg 1\end{array}\right.\label{331}
\end{eqnarray}
where $D_{G}$ is the self-diffusion coefficient.

The main problem now is to find the asymptotic behavior of $\Gamma({\vec
  p},s)$. If $\Gamma({\vec p},s)\propto 1/s^{1-\beta}$ at $s\rightarrow 0$,
where $\beta<1$, one should expect a renormalization of the Rouse dynamics.

Let us consider first the dynamics in the intermediate displacement regime
\begin{eqnarray}
l^{2}<Q(t)<R_{G}^{2}\propto l^{2}{\cal N}^{2/d_{f}^{o}}\label{332}.
\end{eqnarray}
We will show now that the asymptotic behavior of $\Gamma({\vec p},s)$
substantially depends on which limit in eqs.~(\ref{330})-(\ref{331})
is relevant for the dynamics.
\begin{enumerate}
\item Let us assume first that the dominant contribution for the integral over
  the wave vectors in eq.~(\ref{236}) comes from the interval 
\begin{eqnarray}
{\cal N}^{-1}\ll(kl)^{d_{f}^{o}}<1\label{333}.
\end{eqnarray}
Then we can use the second case in eqs.~(\ref{330}) and (\ref{331}) as an input
in the integral (\ref{236}). By performing first the integration over wave
vector ${\bf k}$, and then over ${\vec x}$, one can derive the result
\begin{eqnarray}
{\cal N}\Gamma({\vec
  p},t)=\beta|V({\bf k}={\bf 0})|^{2}\frac{S_{\rm st}({\bf k}={\bf 0})}{l^{d+2}}\cdot\frac{1}{\left(\frac{t}{\tau_{0}}\right)^{\beta}}\label{334}
\end{eqnarray} 
where 
\begin{eqnarray}
\beta=z_{0}(d-d_{\rm uc}+2)\label{335}.
\end{eqnarray}
The Laplace transformation of eq.~(\ref{334}) at $\tau_{0}s\ll1$ is given by
\begin{eqnarray}
{\cal N}\Gamma({\vec
  p},s)=\beta|V({\bf k}={\bf 0})|^{2}\frac{S_{\rm st}({\bf k}={\bf 0})}{l^{d+2}}\tau_{0}^{\beta}\left(\frac{1}{s}\right)^{1-\beta}\label{336}.
\end{eqnarray}
The condition $\beta<1$ ( which is sufficient for the renormalized Rouse
regime) immediately defines the {\it dynamical upper critical dimension}
\begin{eqnarray}
{\tilde d}_{\rm uc}=\frac{4D}{2-D}=2d_{\rm uc}\label{337}
\end{eqnarray}
i.e.the dimension above which the manifold has the simple Rouse behavior, at
$d={\tilde d}_{uc}$ we have the marginal Rouse behavior and only at
$d<{\tilde d}_{uc}$ are the dynamic exponents $z$ and $w$ renormalized. The
dynamical upper critical dimension ${\tilde d}_{{\rm uc}}$ has been discussed first
in [6,7], but the physical interpretation of this was different. It was
asserted in ref.~[6,7] that at $d<2d_{\rm uc}$ (or $d<2d_{f}^{o}$) the strong
entanglement effects become effective (see discussion in Appendix A).

The substitution of eq.~(\ref{336}) in eq.~(\ref{326}) and performing the
inverse Laplace transformation (by making use of the {\it expansion theorem
  } \cite{Doetsch}) yields
\begin{eqnarray}
C({\vec p },t)=C_{\rm st}(p)\sum_{m=0}^{\infty}\frac{\left[-\epsilon
    A\left(\frac{2\pi
        p}{N}\right)^{2}\left(\frac{t}{\tau_{0}}\right)^{\beta}\right]^{m}}{\Gamma(m\beta +1)}\label{338}
\end{eqnarray}
where $A=T\left[|V({\bf k})|^{2}S_{\rm st}({\bf k}={\bf 0})/l^{d+2}\right]^{-1}$ and $\Gamma(x)$
is the gamma-function. The eq.~(\ref{338}) is close to the stretched
exponential form found by MC-simulation \cite{Paul2}.

We should stress that the eq.~(\ref{338}) was actually calculated in the limit
$p/N\rightarrow 0$. That is the reason why we can use it first of all to
comparison with MC-simulation results on the center of mass mean square
displacement. By using eq.~(\ref{338}) in eq.~(\ref{325}) we obtain
\begin{eqnarray}
Q_{\rm cm}(t)=\frac{{\cal D}_{0}}{{\cal
    N}}\cdot\left(\frac{t}{\tau_{0}}\right)^{w}\label{D0}
\end{eqnarray}
where
\begin{eqnarray}
{\cal D}_{0}=\frac{l^{2}\epsilon A}{\Gamma(\beta+1)}\label{D01}
\end{eqnarray}
and
\begin{eqnarray}
w=\beta=z_{0}(d-d_{\rm uc}+2)\label{340}.
\end{eqnarray}
We will compare the dependence (\ref{D0}) in the next Section with some recent MC
findings.

The using of eq.~(\ref{338}) in eq.~(\ref{324}) yields
\begin{eqnarray}
z=z_{0}\beta=z_{0}^{2}(d-d_{\rm uc}+2)\label{339}.
\end{eqnarray}
Since $\beta<1$, the condition $z<z_{0}$ holds, i.e. the interaction with the
matrix slows down the monomer displacement.
\item If we assume now that the main contribution to the integral (\ref{236})
  comes from the small wave vectors 
\begin{eqnarray}
(kl)^{d_{f}^{o}}{\cal N}\ll 1\label{341} 
\end{eqnarray}
then by making use of the small wave vector approximation in eqs.~(\ref{330})
and (\ref{331}) we arrive at 
\begin{eqnarray}
{\cal N}\Gamma({\vec p},s)\propto s^{(d-d_{\rm uc})/2}\label{342}.
\end{eqnarray}
Since $d>d_{\rm uc}$, the simple Rouse behavior in the small wave vector regime
does not change.

Let us finally consider the large displacement regime
\begin{eqnarray}
R_{G}^{2}\ll Q_{\rm cm}(t)\label{343}.
\end{eqnarray}
In this case one should expect simple diffusive behavior
\begin{eqnarray}
Q_{\rm cm}(t)=dD_{G}t\label{344}
\end{eqnarray}
where the self-diffusion coefficient \cite{3,4,6}
\begin{eqnarray}
D_{G}=\frac{T}{{\cal N}\left[\xi_{0}+{\cal N}\Gamma(p=0,s=0)\right]}\label{345}.
\end{eqnarray}
When the dynamics of the test manifold is characterized by the self-diffusion
coefficient $D_{G}$, the dynamics of the matrix is driven mainly by the
cooperative diffusion coefficient $D_{\rm coop}$ (see eqs.~(\ref{330})-(\ref{331})).
Again, the small wave vectors interval, eq.~(\ref{341}), is relevant. Since in
any case
$D_{\rm coop}\gg D_{G}$, the calculations yields 
\begin{eqnarray}
{\cal N}\Gamma(p=0,s=0)\propto \left(D_{\rm coop}\right)^{-1}{\cal
  N}^{\left(1-\frac{d}{d_{f}^{o}}\right)}\label{346}.
\end{eqnarray}
But $D_{\rm coop}={\cal O}\left(N^{0}\right)$ and $d_{f}^{o}<d$, therefore
$\Gamma(p=0,s=0)\rightarrow 0$ at ${\cal N}\rightarrow \infty$. As a result
$D_{G}=T/{\cal N}\xi_{0}$ and again the simple Rouse behavior is not
renormalized.
\end{enumerate}
\section{Discussion and conclusions}
In the present work, we have derived the GRE (\ref{235})-(\ref{239}) for the
test polymeric manifold in a melt composed of chains of the same nature. In order to calculate it, we have
integrated over the melt's collective variables in the framework of the RPA
and have used the selfconsistent Hartree approximation for the resulting effective action
functional. It is very important that in this GRE the static and dynamic parts
are treated in the same manner.

In particular, if instead of the RPA the mode-coupling approximation (MCA)
would be used for the collective variables (as it was done for linear polymer chains
in \cite{3,4}), one should substitute the RPA-correlators in eqs.~(\ref{236})
and(\ref{239}) with the {\it full} correlators. It is then not obvious,
how e.g. the simple screened form (\ref{240}) for the effective interaction potential
could be obtained. In this respect generally speaking the dissipative and the
reactive parts of the GRE in \cite{3,4} do not conform.

The GRE derived here (as well as the GRE from \cite{3,4,4',4'',5,6}) cannot describe the
reptational dynamics or entanglements. In order to do this, one should
incorporate topological constraints in the microscopic equation of
motion. The extensive polymer mode-coupling approach (PMCA) [4-7] which was
worked out to explain the entanglement dynamics (without the reptational model
[16,17]) is, in our opinion, the result of misinterpretation of the GRE (see
Appendix A).  

We have shown that at $d>d_{\rm uc}$ the excluded volume interaction is screened
out and the manifold is Gaussian with the exponent $\zeta_{0}=(2-D)/2$
\cite{18,19}. Nevertheless the dynamic behavior is renormalized whenever
$d_{\rm uc}<d<{\tilde d}_{uc}$, where ${\tilde d}_{uc}$, the {\it dynamic upper
  critical dimension}, is given by eq.~(\ref{337}), i.e. the reactive and the
dissipative forces do not screen out simultaneously.

For example, for the melt of polymer chains $(D=1)$ $d_{\rm uc}=2$ and ${\tilde
  d}_{uc}=4$, and in the 3 - dimensional space one should expect the Gaussian
static behavior but renormalized-Rousian dynamics. According
to eqs.~(\ref{340})-(\ref{339}) at $d=3$ we obtain the exponents 
\begin{eqnarray}
w=\beta=3/4=0.75
\label{89}
\end{eqnarray}
and
\begin{eqnarray}
2z=3/8=0.375\label{90}.
\end{eqnarray}
In eq.~(\ref{338}) we can take with a good
approximation $\Gamma(m\beta+1)\approx \Gamma(m+1)$, then the Rouse modes
correlator has the stretched exponential form
\begin{eqnarray}
C({\vec p},t)=C_{\rm st}(p)\exp\left\{-\epsilon A\left(\frac{2\pi
      p}{N}\right)^{2}\left(\frac{t}{\tau_{0}}\right)^{\beta}\right\}\label{41}.
\end{eqnarray}
The eqs.~(91-93) were derived first in [4] as the solution of the first order
approximation (so called RR-model). According to ref.~[4-6] this regime describes
the onset of the entanglenment dynamics (see Appendix A).

As we have mentioned above eqs.~(\ref{334}), (\ref{336}) and (\ref{338}) which lead
to eq.~(\ref{41}) was actually calculated at $p\rightarrow 0$. That means that
an experiment or simulation on the center of mass mean square displacement,
$Q_{\rm cm}(t)$, is the best candidate for the comparison with our results.

MC-simulations of the bond fluctuation model \cite{Paul} as well as the
MD-simulations \cite{Smith,Kopf} of the athermal polymer melt have been
undertaken. Recently also the static and dynamic properties of a realistic
polyethylene melt have been studied by the extensive MD-simulations
\cite{Paul1,Paul2}. Both in MC and MD simulations [24-28] a slowed-down motion at
intermediate times for the center of mass mean square displacement is clearly
observable. It was found e.g. that for the
chain length $N=200$ at the relatively short time ($t\le 3\cdot 10^{6}$ MCS in
\cite{Paul}) $Q_{\rm cm}(t)\propto t^{w}$ with $w=0.8$ (instead of $w=1$) in
\cite{Paul} and $w=0.71$ in \cite{Smith}. This corresponds to our prediction
$w=0.75$. At larger times and scales the crossover to the reptational (for the
long noncrossable chains) regime can be seen. This deviation from the simple
Rouse regime also occurs for very short chains, $N=20$, which clearly are not
entangled. The regime $Q_{\rm cm}(t)\propto t^{0.8}$ for the short-time regime has
actually been observed first by Kremer and Grest \cite{15}.
 
In Fig.~1 we have summarized the overall schematic behaviour for $Q_{\rm
  cm}(t)$. At the relatively short times, $\tau_{0}<t\ll\tau_{R}$, and displacements,
$l^{2}<Q_{\rm cm}(t)\ll R_{G}^{2}$, the test chain dynamics is mainly ruled by
fluctuations from the interval (\ref{333}), i.e. the Rouse dynamics is
renormalized with the exponent (\ref{89}). The picture which underlies this
renormalization is visually represented in Fig.~2. The diffusing test chain
in this case experiences mainly the short wavelength density fluctuations of
the melt and, as a result, it is weakly "pinned" on the "lattice" induced by
the density fluctuations. This "pinning" naturally results in a subdiffusive
$(w<1)$ behavior at $\tau_{0}<t\ll \tau_{R}$.

In the opposite limit, $\tau_{R}\ll t$ and $R_{G}^{2}\ll Q_{\rm cm}(t)$, the long
wavelength fluctuations from the interval (\ref{341}) are relevant. Then the
picture of the interplay between the test chain and the melt density fluctuations
is given in Fig.~3. In this limit the melt almost does not influence the test
chain and the simple Rouse regime is recovered.

The crossover area, $t\approx \tau_{R}$ and $Q_{\rm cm}(t)\approx
R_{G}^{2}$, is not amenable for the theoretical investigation mainly because
of the lack of complete analytical expressions for the correlators
(\ref{330})-(\ref{331}). Note that the renormalized curve in Fig.~1 converges
asymptotically to the simple Rouse curve from above. This is assured by the
relationship ${\cal D}_{0}>dT\tau_{0}/\xi_{0}$, where the renormalized beads
diffusion coefficient, ${\cal D}_{0}$, is given by eq.~(\ref{D01}). This
condition can be represented in the form
\begin{eqnarray}
\frac{|V({\bf k})|^{2}}{T^{2}}\cdot\frac{S({\bf k}={\bf 0})}{dl^{d}}<1\label{cond}.
\end{eqnarray}
In \cite{Paul} it has been explicitly shown that the renormalized beads
diffusion coefficient (it is called an acceptance rate in \cite{Paul}) is
larger than its Rouseian counterpart, i.e. the condition (\ref{cond}) holds for
the real systems. The renormalized Rouse regime which is given in Fig.~1 is
qualitatively the same as in MC [24] and MD [25-27] simulations (see
e.g. Fig.~8b in [24], Fig.~9 in [27] and Fig.~3 in [28]).

The slight stretching of the Rouse modes time correlation function has been
found in [27,28] for modes with $p=1,2,3$ (which still satiesfy the static
$p^{-2}$-law). This is different from ref.~[26] where no deviation from the
ideal Rouse behavior for the higher modes has been observed. So as a future
perspective it would be very interesting to solve eq.~(44) numerically and
compare the effects of the renormalization at finite $p/N$ with the results of
simulations.

In [29,30] the MC-simulations for the dynamics of the athermal polymer melt
have been undertaken. Special attention has beeb paid to the comparison of the
crossable and noncrossable chains. These types of simulation are specially
``designed'' to check our results: The chains are long enough to follow the
Rouse dynamics and to study its renormalization by interactions. It was shown
in [29,30] that at relatively short wavelength Rouse modes, $N/p\le 6$ and
$N=500$, the stretching parameter in the Rouse mode correlation function
$\beta\approx 0.8$. It is probably the result of the lattice structure
structure influence. Unfortunately it is not clear why the static correlator
of the Rouse modes deviates from the ideal values even for the long
wavelength modes (see e.g. Fig.~3 in [30]). For the relatively long wavelength
Rouse modes, $p/N\ge 10$ and $N=500$ (for crossable chains), the simple Rouse
dynamics holds (i.e. $\beta=1$ see Fig.~9 in [30]), but since the plot of
$Q_{\rm cm}(t)$ is not given explicitly, it stays inclear how the mode
$p\rightarrow 0$ is renormalized. 

The {\it temperature dependence} of the renormalized Rouse regime is
determined by the excluded volume interaction potential $V({\bf k})$. It is
different from the static, where e.g. the screening effect does not depend
on temperature at any potential. The condition for the occurence of
renormalization is given by $\xi_{0}\ll \int_{0}^{t_{max}}{\cal N}\Gamma({\vec
  p},t)dt$, which, after using eq.~(\ref{334}), takes the form  
\begin{eqnarray}
1\ll
\frac{|V({\bf k})=0|^{2}}{T^{2}}\cdot\frac{S_{\rm st}({\bf k}={\bf 0})}{d(1-\beta)l^{d}}\left(\frac{t_{max}}{\tau_{0}}\right)^{1-\beta}\label{42}
\end{eqnarray}
where $t_{max}\propto\tau_{R}\approx\tau_{0}N^{2}$. For the athermal case one
can take as an estimate $V({\bf k}={\bf 0})\approx Tv$ (Edwards potential), and
the condition (\ref{42}) always holds. For the pseudo-potential $V\approx
\varepsilon v$, where $\varepsilon$ has the dimension of molecular
energy, the condition (\ref{42}) is violated at rather high temperatures and
the dynamics becomes Rousian.      
\begin{acknowledgments}
We are grateful to J. Bashnagel, K. Binder, K. Kremer, B. D\"unweg,
K. M\"uller-Nedebock, T. Liverpool and W. Paul for
interesting and useful discussions. The authors thank the Deutsche
Forschungsgemeinschaft (DFG), the Sonderforschungsbereich SFB 262 and the
Bundesministerium f\"ur Bildung und Forschung (BMBF) for financial support of
the work.
\end{acknowledgments}
\begin{appendix}
\section{Comparison with Schweizer's polymer mode-coupling approach (PMCA)}
The purpose of this Appendix is to critically analyze the basic aspects of PMCA
\cite{4,4',4'',5} and to show that the pseudo-reptational exponents, which were
obtained in PMCA, are a result of misinterpretation of the GRE.

In Schweizer's PMCA the projection operator formalism and the
mode-coupling approximation were used in order to derive a GRE for the Gaussian
chain. As a result the efffects of interaction of the test chain with the
other chains are present in the form of a memory function. The resulting GRE
for the test chain correlation function $C(\alpha,\beta;t)=\left<{\bf
    R}(\alpha,t){\bf R}(\beta,0)\right>$ takes the form \footnote{We use here
  the notation of ref.~\cite{4}}
\begin{eqnarray}
\xi_{0}\frac{\partial}{\partial
  t}C(\alpha,\beta;t)+\int_{0}^{N}d\gamma\int_{0}^{t}dt'\Gamma(\alpha,\gamma;t-t')\frac{\partial}{\partial t'}C(\gamma,\beta;t')=K_{s}\frac{\partial^{2}}{\partial \alpha^{2}}C(\alpha,\beta;t)\label{B1}
\end{eqnarray}
where $K_{s}=3k_{B}T/\sigma^{2}$ is the entropic spring constant and the
memory function
\begin{multline}
\Gamma(\alpha,\beta;t)=\left(\frac{k_{b}T}{2\pi^{2}}\right)\int_{0}^{\sigma^{-1}}dk\:k^{4}{\hat
  C}^{2}({\bf k}){\hat \omega}^{2}({\bf k})\rho_{m}{\hat S}({\bf k}){\hat F}_{s}^{Q}({\bf
  k},t)\\
\times\int_{0}^{N}d\gamma d\delta\:{\hat \omega}^{-1}(\alpha,\beta;{\bf k}){\hat
  \omega}^{Q}(\gamma,\delta;{\bf k},t){\hat \omega}^{-1}(\delta,\beta;{\bf k})\label{B2}
\end{multline}
where $\rho_{m}$ is the average density, ${\hat C}({\bf k})$ is the direct
correlation function, ${\hat \omega}({\bf k})$ and ${\hat S}({\bf k})$ are static
correlators for the test chain and matrix correpondingly, ${\hat
  \omega}^{Q}(\alpha,\beta;{\bf k},t)$ is the dynamic test chain density correlator
associated with segments $\alpha$  and $\beta$, ${\hat
  \omega}(\alpha,\beta;{\bf k})={\hat \omega}^{Q}(\alpha,\beta;{\bf k},t=0)$ and ${\hat
  F}_{s}^{Q}({\bf k},t)$ is the normalized dynamical collective matrix
correlator. As usual the superscript $Q$ denotes the evolution via projected
dynamics. 

As a main approximation the projected single chain
dynamics was substituted by so-called "renormalized Rouse" (RR) dynamics
\begin{eqnarray}
{\hat \omega}^{Q}(\alpha,\beta;{\bf k},t)\rightarrow{\hat
  \omega}^{RR}(\alpha,\beta;{\bf k},t)\approx{\hat \omega}(\alpha,\beta;{\bf k}){\hat
  F}^{RR}(k,t)\label{B3}.
\end{eqnarray}
This RR-dynamics is a first order approximation of an iterative solution
of eq.~(\ref{B1}) when as a zeroth order approximation the well-known Rouse
(R) expression 
\begin{eqnarray} 
{\hat \omega}^{Q}(\alpha,\beta;{\bf k},t)\rightarrow{\hat
  \omega}^{R}(\alpha,\beta;{\bf k},t)\approx{\hat \omega}(\alpha,\beta;{\bf k}){\hat
  F}^{R}(k,t)\label{B4}
\end{eqnarray}
is employed. The corresponding justification given in \cite{4,4',4''} to take the
$RR$-approximation as a starting point "in crude analogy with Enskog theory"
remains questionable.

For the projected collective correlation the real (not RR) dynamic evolution
is employed
\begin{eqnarray}
{\hat F}_{s}^{Q}(k,t)\rightarrow{\hat F}_{s}(k,t)\label{B5}
\end{eqnarray}
This approximation could be justified for small ${\bf k}$ \cite{Forster} and was
used e.g. in the glass transition theory \cite{Goetze}.

The main problem with the present analysis of eq.~(\ref{B1}) is that it cannot
be solved iteratively. It is easy to see that eq.~(\ref{B1}) is
substantially non-linear because in the memory function (\ref{B2}) the dynamic
correlator is given by
\begin{eqnarray}
\omega^{Q}(\alpha,\beta;{\bf k},t)&=&\left<\exp\left\{-i{\bf k}\left[{\bf
        R}(\alpha,t)-{\bf R}(\beta,0)\right]\right\}\right>\nonumber\\
&=&\exp\left\{-\frac{k^{2}}{3}\left[C(\alpha,\alpha;t=0)-C(\alpha,\beta;t)\right]\right\}\label{B6}.
\end{eqnarray}
The first line in eq.~(\ref{B6}) implies that the real dynamical evolution is
used and the second equality comes from the fact that the fluctuations of a
$R$-variable are Gaussian. As a result eq.~(\ref{B1}) is substantially
non-linear and should be treated self-consistently around the bifurcation
points as in the glass transition theory \cite{Goetze}. Substantially
non-linear means that the range of parameters (temperatures, length of
chains, time, etc.) is such that the mode-coupling friction term in
eq.~(\ref{B1}) is comparable to or much larger than the bare friction one. But this
is exactly the range which PMCA investigates as a crossover from Rouse
to entangled dynamics.

Instead of selfconsistent solutions around  a bifurcation point iterations are
actually used. As a zeroth order approximation this yields the $R$-expression,
and the first order approximation - the $RR$-model - is used as a starting
point for the description of entanglements. The second order approximation
gives already the result which looks like entangled dynamics. But one can see that this
iteration procedure is divergent as it should be for a substantially
non-linear equation.

For example, the Rouse friction coefficient $\xi_{R}\propto D_{R}^{-1}\propto
N$, the $RR$-model leads to $\xi_{RR}\propto D_{RR}^{-1}\propto N^{3/2}$ and
at last the next approximation yields $\xi_{\rm coil}\propto D_{\rm coil}^{-1}\propto
N^{2}$.

There is a number of differences between Schweizer's GRE and the GRE which was
derived here. First of all, the statics and dynamics of the matrix (or melt)
in eqs.~(35)-(41) are described by RPA-correlators. This results in the
conventional static screening (see eq.~(40)). In order to introduce full
correlators we should go beyond the Gaussian approximation in the expansion
(17), i.e. take into account the melt fluctuations around the RPA [34]. This
is a tough problem even for statics and it is not clear beforehand whether the
simple bilinear form of the memory function (36) will remain. Unlike this, in
Schweizer's GRE [4] the {\it full} melt correlator appears in the memory
function whereas for the static part it is simply apriori taken to be the screened
form of the interaction (see eq.~(2.3) in ref.~[4]).

Secondly, by the analytical investigation of Schweizer's GRE different
analytically tractable formulas for the matrix correlator were assumed. For
example in sec.~III c of ref.~[4] ("Center of mass diffusion and shear
viscosity") or in sec.~III in ref.~[5] the frozen matrix
i.e. ${\hat F}_{s}({\bf k},t)=1$ was adopted. This brings the result for the RR-model $D_{G}\propto N^{-3/2}$, which was mentioned above. In our GRE we cannot
consider this case because in RPA-approximation the matrix is not frozen. But
in most of the computer simulations [24-29] the athermic polymer melts have
been studied, which cannot be frozen. In this case at small wave vectors
$((kl)^{d_{f}^{o}}{\cal N}\ll 1)$ the dynamics of the matrix is driven by
$D_{\rm coop}$ and the simple Rouse dynamics of the test chain is recovered.

For the small or intermediate displacements, $l^{2}<Q_{{\rm
    cm}}(t)<R_{G}^{2}$, the so called Vineyard approximation, i.e. ${\hat
  F}_{s}({\bf k},t)={\hat F}({\bf k},t)$, was adopted. This leads for the
RR-model (first order approximation) to formally the same mathematics as in our
sec.~III B(i). As a result the values of the exponents $w$ and $z$ (see
eqs.~(91)-(92)) as well as the dynamical uppper critical dimension, ${\tilde
  d}_{\rm uc}=2d_{f}^{o}$, are the same. This value of ${\tilde d}_{\rm uc}$
is assured by the short range nature of the interaction function $V({\bf r})$
in the memory kernel which leads to the scaling for the number of {\it
  dynamically} effective contacts: $Z_{dyn}\propto N^{2-d/d_{f}^{o}}$. This
scaling law has been discussed first in ref.~[6,7]. Nevertheless the
criterion $d<2d_{f}^{o}$ has been considered as a necessary condition for the
onset of the entangled dynamics. As we have already mentioned above in order
to obtain the entangled dynamics exponents in PMCA the next (second order)
iteration is implemented. In the spirit of PMCA we could use e.g. the
renormalized Rouse exponents as an input in the r.h.s. of eqs.~(83)-(84) and
rederive the pseudo-reptational exponents
\begin{eqnarray}
w_{\rm rep}=\beta_{\rm rep}=z(d-d_{\rm uc}+2)=\frac{3}{16}\cdot 3=\frac{9}{16}\label{B8}
\end{eqnarray}
and
\begin{eqnarray}
2z_{\rm rep}=2z_{0}\beta_{\rm rep}=\frac{1}{2}\cdot\frac{9}{16}=\frac{9}{32}\label{B9}.
\end{eqnarray}
These exponents have been given in \cite{4''}. Unfortunately, this way of analysis does not seem to bring reliable results.

In the present work we have made every effort to prove that the excluded
volume interaction results in the renormalized Rouse regime in the melt. We
believe that the topological constraints are essential for the entangled
dynamics.
The recent MC-simulation \cite{20,20'} have shown the decisive role of the
topological constraints underlying the reptational dynamics. In particular the
crossable and noncrossable models of rather long chains ($10<N<500$) were used
for the simulation of statics as well as dynamical properties of polymer
melts. It was shown that the static properties of both models are absolutely
identical (see e.g. Fig.~3 in \cite{20}). On the other hand the dynamic
behavior is completely different. The crossable chains (irrespective of short
or long) show at the relatively long wavelengths the simple Rouse behavior
(see e.g. Fig.~6 and Fig.~9 in \cite{20'}). Also the
self-diffusion coefficient has the Rouseian scaling $D_{G}\propto N^{-1}$
(Fig.~(7) in \cite{20}). For the noncrossable
chains (at $N=100, 300$ and $500$) the stretched exponential regime is found
at all mode wavelengths. The self-diffusion coefficient scales as
$D_{G}\propto N^{-2.08}$ at $N>40$ and the mean-square displacement
$Q(t)\propto t^{1/4}$ in the corresponding time window.

This proves that the chain-crossing condition does not touch the static
properties but has a dramatic effect on the dynamics. On the other hand, the
static input completely predestines the dynamic behavior in the
PMCA-formalism. This contradiction unfortunately put severe doubt on the
PMCA. We rather feel that the explicit taking into account of topological
constraints in the microscopical equations of motions is absolutely required. 
\end{appendix}

\newpage
\begin{center}
\begin{figure}
\epsfig{file=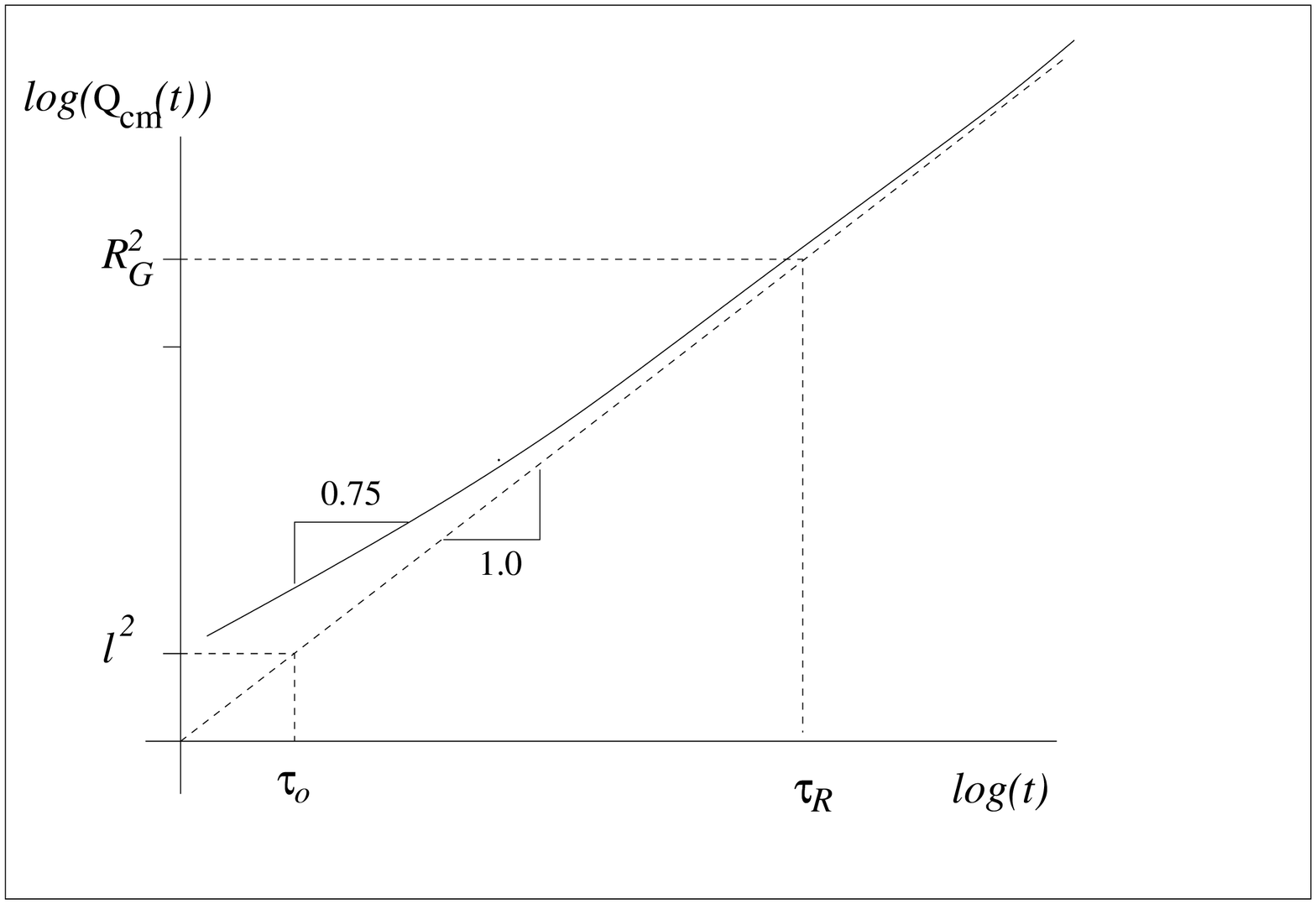,width=10cm,height=6cm}
\caption{A schematic plot of $Q_{\rm cm}(t)$ for the simple Rouse (dashed line)
  and the renormalized Rouse (solid line) dynamics}
\end{figure} 
%\newpage
%\begin{center}
\begin{figure}
\epsfig{file=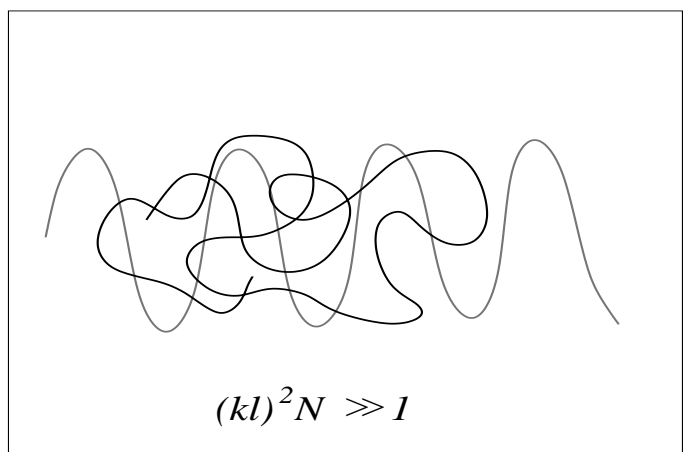,width=9cm, height=6cm}
\caption{The test chain experiences the short wavelength melt density
  fluctuations and as a result it is weakly "pinned".}
\end{figure} 
%\newpage
%\begin{figure}
%\epsfig{file=Fig2bn.eps,width=9cm, height=6cm}
%\caption{Intermediate scales}
%\end{figure}
%\newpage
\begin{figure}
\epsfig{file=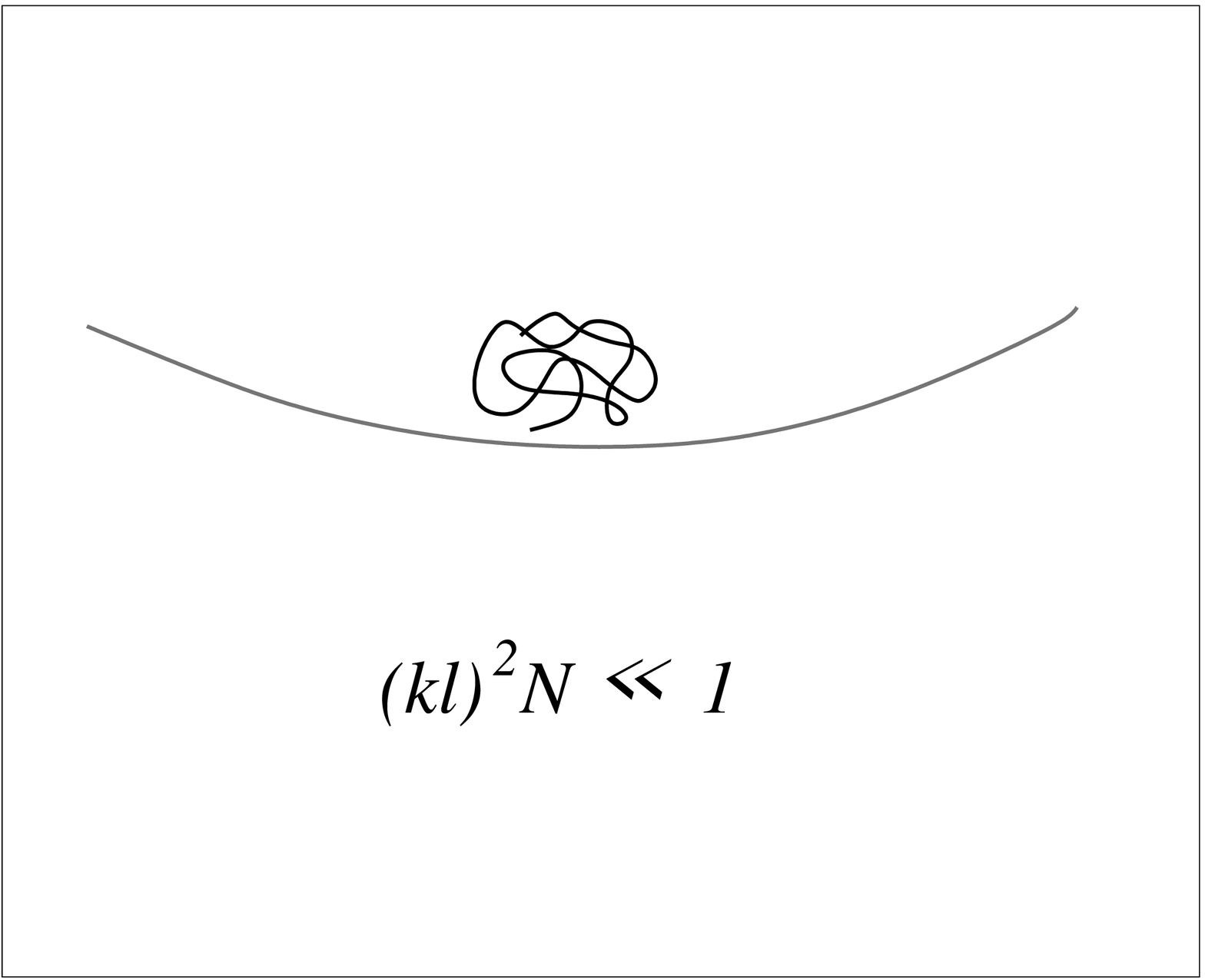,width=9cm, height=6cm}
\caption{The interplay of the test chain and very long wavelength density
  fluctuations does not influence the Rouseian dynamics.}
\end{figure}
\end{center}
\end{document}